\documentclass[aps,prb,twocolumn,showpacs,preprintnumbers,amsmath,amssymb]{revtex4-1}

\usepackage{graphicx}
\usepackage{dcolumn}
\usepackage{bm}
\usepackage{color}

\usepackage{ulem}

\newcommand{\ket}[1]{|#1\rangle}

\newcommand{\be}{\begin{equation}}
\newcommand{\ee}{\end{equation}}
\newcommand{\bea}{\begin{eqnarray}}
\newcommand{\eea}{\end{eqnarray}}

\newcommand{\eq}[1]{Eq.~(\ref{#1})}
\newcommand{\fig}[1]{Fig.~\ref{#1}}

\newcommand{\e}{\varepsilon}

\newcommand{\s}{\sigma}
\newcommand{\G}{\Gamma}

\begin{document}

\title{Superconducting proximity effect and zero-bias anomaly in transport through
quantum dots weakly attached to ferromagnetic leads}

\author{Ireneusz Weymann}
\email{weymann@amu.edu.pl}
\affiliation{Faculty of Physics, Adam
Mickiewicz University, 61-614 Pozna\'n, Poland}

\author{Piotr Trocha}
\affiliation{Faculty of Physics, Adam
Mickiewicz University, 61-614 Pozna\'n, Poland}

\date{\today}

\begin{abstract}
The Andreev transport through a quantum dot coupled to two external 
ferromagnetic leads and one superconducting lead is studied theoretically
by means of the real-time diagrammatic technique in the sequential and cotunneling regimes.
We show that the tunnel magnetoresistance (TMR) of the Andreev current displays
a nontrivial dependence on the bias voltage and the level detuning, and can be described
by analytical formulas in the zero temperature limit. The cotunneling processes
lead to a strong modification of the TMR, which is most visible
in the Coulomb blockade regime.
We find a zero-bias anomaly of the Andreev differential
conductance in the parallel configuration, which is associated with a nonequilibrium
spin accumulation in the dot triggered by Andreev processes.
\end{abstract}

\maketitle


\section{Introduction}


The large tunability of quantum dot properties by applying proper gate voltages
makes these structures very promising for spintronic and quantum information applications.
\cite{awschalom_book,zutic04}
These nanoscale structures enable the observation of various novel phenomena,
and exhibit effects known from solid state physics, atomic physics or quantum optics.
\cite{brandesPR05}
Moreover, nanoscopic systems with quantum dots coupled to
ferromagnetic leads can exhibit a considerable tunnel
magnetoresistance (TMR) effect and can be used for spin current generation.
\cite{sahoo,hamayaAPL08,barnasJPCM08,birk,csonka12}
The theoretical and experimental investigations of transport
through such structures are thus of great current interest.
Transport properties of single quantum dots attached to ferromagnetic leads
have been extensively investigated both experimentally
\cite{pasupathy,mentel,hamaya1,hamaya2,hamaya3,yang,hamaya4,gaass}
and theoretically.~\cite{rudzinski,braun,cottet,weymann}
The physics however becomes much more interesting and fascinating
when these nanostructures are in proximity with a superconductor.
In such hybrid structures, transport properties are determined
by the interplay of the spin-dependent tunneling and superconducting correlations.~\cite{csonka}

The hybrid nanoscopic structures consisting of quantum dots
attached to normal and superconducting external leads have attracted great interest,
mainly due to the possibility of creation of nonlocal entangled electron pairs.~\cite{hofstetterSC,herrmann10SC}
The generation of entangled pairs is strictly connected with the
processes known as crossed Andreev reflections (CARs).
In contrast to the usual Andreev reflection, in CAR the hole is reflected back into 
the electrode, which is spatially separated from the lead, from which the incoming electron arrives.
Such processes have been investigated both theoretically and
experimentally in hybrid metallic structures.~\cite{deutscherAPL00,beckmannPRL04,russoPRL05}
Moreover, very recently more efficient and easy tunable Cooper pair beam splitter
has been implemented by using double quantum dot structures.~\cite{hofstetterSC,herrmann10SC}
Quantum dots provide thus an interesting and promising route for
manipulating entangled electrons by electrical means in a controllable fashion,
which with no doubt is of great importance for quantum computation and quantum information.
\cite{awschalom_book}
However, to exploit them efficiently and reliably, it is crucial to understand
various properties of such hybrid nanostructures,
including transport properties due to Andreev reflection.

In this paper, we therefore investigate the Andreev transport through single-level
quantum dots with a special focus on the interplay
of the spin-dependent and Andreev tunneling processes.
For this, we assume that the dot is coupled to two ferromagnetic leads
and one s-wave superconducting electrode, where the coupling to ferromagnets
is weak, while the coupling with the superconductor can be arbitrary.
We assume that the magnetic moments of the ferromagnetic leads
may be aligned either in parallel or antiparallel. The difference in these two magnetic configurations
gives rise to the tunnel magnetoresistance (TMR) of the Andreev current.
Since by changing the magnetic configuration the
amount of CAR is affected, by studying the behavior
of the TMR one can obtain information about the role of crossed
Andreev processes in transport.
We note that transport properties of hybrid systems consisting of quantum dot coupled
to ferromagnetic and superconducting electrodes have already been studied.
The considerations however concerned the case of one
ferromagnetic and one superconducting lead.~\cite{feng03,cao,pengZhang,csonka,wysokinskiJPCM}
In such geometry, the crossed Andreev reflections are not allowed.
On the other hand, CAR was studied in the case of two normal leads
and one superconducting lead in the absence of intra-dot Coulomb interactions.
\cite{sunSC01,golubev07}
Nevertheless, the Coulomb interactions, which are strong in typical quantum dots,
may weaken the proximity effect induced on the dot
or even completely destroy it.~\cite{konig09,konigPRB10}
Thus, the effects of Coulomb correlations should be regarded
on an equal footing with the other effects, such as e.g. superconducting correlations,
nonequilibrium or the magnetism of the leads.
Moreover, the case when the normal leads are replaced
by ferromagnets was studied theoretically however only
in the sequential tunneling regime.~\cite{konig09,konigPRB10}
The sequential tunneling approximation may however lead to wrong predictions,
especially for the TMR in the Coulomb blockade regime.~\cite{weymann}
The goal of the present work is thus to extend theses studies,
by calculating the Andreev transport for realistic quantum dot parameters,
including both the sequential and cotunneling processes.

To calculate the basic transport characteristics, both in equilibrium
and nonequilibrium, we employ the real-time diagrammatic technique.~\cite{schoeler}
Particularly, by taking into account the first and second-order
diagrams, we calculate the Andreev current,
differential conductance and the tunnel magnetoresistance.
We discuss the influence of cotunneling processes
on the sequential Andreev transport,
which has been previously studied in Refs.~[\onlinecite{konig09}] and [\onlinecite{konigPRB10}],
and show that it leads to a strong modification of the TMR
for bias voltages where the Andreev current is suppressed due to the charging effects.
Moreover, we predict a zero-bias anomaly in the differential conductance
of Andreev current for the parallel magnetic configuration of the device,
which is due to the nonequilibrium spin accumulation in the dot
induced by cotunneling processes.

The paper is organized as follows.
In Sec. II we describe the model and method used in calculations.
Section III is devoted to numerical results and their discussion.
We first analyze the dependence of the Andreev current
and differential conductance on the bias voltage and the detuning
of the dot level and then discuss the behavior of the associated TMR.
We also analyze the behavior of the zero-bias
anomaly on various model parameters.
Finally, the conclusions are given in Sec. IV.


\section{Theoretical description}


\begin{figure}[t]
\includegraphics[width=0.7\columnwidth]{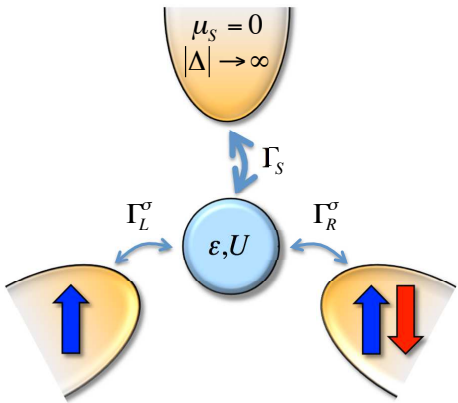}
  \caption{ \label{Fig:1}
  (color online) The schematic of a quantum dot strongly
  coupled to an s-wave superconductor
  and weakly coupled to two ferromagnetic leads.
  The coupling to superconductor is described by $\G_S$,
  while the spin-dependent couplings to ferromagnetic leads
  are denoted by $\G_{L}^\sigma$ and $\G_{R}^\sigma$, respectively.
  The magnetizations of the ferromagnets can form
  either the parallel or antiparallel magnetic configuration, as indicated.
  The superconducting gap $\Delta$ is assumed to be the largest
  energy scale in the problem, $|\Delta|\to\infty$, and the electrochemical potential
  of superconducting lead is set to zero, $\mu_S = 0$.
  $\varepsilon$ denotes the dot level energy
  and $U$ is the Coulomb repulsion on the dot.}
\end{figure}

\subsection{Model Hamiltonian}


We consider a single-level quantum dot tunnel coupled to two
ferromagnetic leads and one superconducting electrode,
as shown schematically in Fig.~\ref{Fig:1}.
The magnetizations of the leads are assumed to be
collinear, and the system can be either in the parallel or antiparallel
magnetic configuration. The magnetic configuration 
of the system can be switched by applying a weak external magnetic field.
Since there is also superconducting electrode attached to the dot,
the magnetic field that switches the orientations of the magnetizations of ferromagnetic
leads needs to be smaller than the corresponding critical magnetic field of the superconductor.
The Hamiltonian of the system acquires the following form
\begin{equation}\label{Eq:1}
  H=\sum_{\beta=L,R}H_{\beta} + H_S +H_{QD}+H_T,
\end{equation}
where the first term, $H_{\beta}$, describes the left ($\beta=L$) and
right ($\beta=R$) electrodes in the noninteracting quasiparticle
approximation,
$H_{\beta}=\sum_{\textbf{k}\sigma}\varepsilon_{\textbf{k}\beta\sigma}
c_{\textbf{k}\beta\sigma}^{\dagger}c_{\textbf{k}\beta\sigma}$.
Here, $c_{\textbf{k}\beta\sigma}^{\dagger}$ ($c_{\textbf{k}\beta\sigma}$)
is the creation (annihilation) operator of an electron
with the wave vector $\mathbf{k}$ and spin $\sigma$ in the lead $\beta$,
whereas $\varepsilon_{\textbf{k}\beta\sigma}$ denotes
the corresponding single-particle energy.
The second term in Eq.~(\ref{Eq:1}) describes
the s-wave BCS superconducting lead in the mean field approximation
\begin{eqnarray}\label{Eq:2}
H_{S}&=&\sum_{{\mathbf
k}}\limits\sum_\sigma\limits\varepsilon_{{\mathbf k}S\sigma}
     c^\dag_{{\mathbf k}S\sigma}c_{{\mathbf k}S\sigma}
     \nonumber\\
     &+&
     \sum_{{\mathbf k}}\limits\left(\Delta^{\ast} c_{{\mathbf
k}S\downarrow}c_{{-\mathbf k}S\uparrow}+\Delta c^{\dag}_{{-\mathbf
k}S\uparrow}c^{\dag}_{{\mathbf k}S\downarrow}\right),
\end{eqnarray}
with $\varepsilon_{{\mathbf k}S \sigma}$ denoting the relevant
single-particle energy and $\Delta$ being the order parameter
of the superconductor. Without loss of generality,
the order parameter can be chosen real and positive, $\Delta=|\Delta|$.
The third term of the Hamiltonian (\ref{Eq:1}) describes the single-level quantum dot
and has the following form
\begin{equation}\label{Eq:3}
  H_{QD}=\sum_{\sigma}\varepsilon d_{\sigma}^{\dagger}d_{\sigma}+Un_{\uparrow}n_{\downarrow},
\end{equation}
where $\varepsilon$ denotes the dot's level energy 
and $U$ is the corresponding Coulomb repulsion energy.

Finally, the tunneling of electrons between all the leads ($L,R,S$)
and the quantum dot can be modeled by the Hamiltonian
\begin{equation}\label{Eq:4}
H_T=\sum_{\mathbf{k}\sigma}\limits\sum_{\beta=L,R,S}
   \limits (V_{\mathbf{k}\sigma}^\beta c^\dag_{\mathbf{k}\beta\sigma}d_{\sigma}+\rm
   h.c.),
\end{equation}
with $V_{\mathbf{k}\sigma}^\beta$ denoting the relevant tunneling matrix elements.
In the following we assume that these matrix elements
are $\mathbf{k}$ and $\sigma$ independent.
The coupling of the dot to the ferromagnetic lead ($\beta=L,R$)
can be parametrized by $\Gamma_{\beta}^{\sigma} = 2\pi|V^{\beta}|^2\rho_{\beta}^\sigma$,
where $\rho_{\beta}^\sigma$ is the density of states of lead $\beta$
for spin $\sigma$.
Within the wide band approximation the couplings become
energy-independent and constant.
By introducing the spin polarization of lead $\beta$,
$p_\beta=(\rho_{\beta}^+-\rho_{\beta}^-)/(\rho_{\beta}^++\rho_{\beta}^-)$,
the couplings can be written in the form,
$\Gamma_{L}^{\sigma} = \Gamma_L (1+\tilde{\sigma}p)$ and
$\Gamma_{R}^{\sigma} = \Gamma_R(1\pm\tilde{\sigma}p)$,
with $\tilde{\sigma}=1$ for $\sigma=\uparrow$, and
$\tilde{\sigma}=-1$ for $\sigma=\downarrow$, and the upper (lower)
sign in $\Gamma_{R}^{\sigma}$ corresponding to the parallel
(antiparallel) magnetic configuration.

As we are interested in the Andreev transport regime,
in our considerations we can take the limit of
an infinite superconducting gap, $\Delta\rightarrow\infty$.
Then, the quantum dot coupled to the superconducting lead
can be described by the following effective Hamiltonian~\cite{rozhkov}
\begin{equation}\label{Eq:5}
  H_{QD}^{\rm eff}=H_{QD}-
  \frac{\Gamma_S}{2}d_{\uparrow}^{\dagger}d_{\downarrow}^{\dagger}- \frac{\Gamma_S}{2}d_{\downarrow}d_{\uparrow}\,.
\end{equation}
It can be clearly seen that the superconducting proximity effects
are included in the last two terms of Eq.~(\ref{Eq:5}),
where the effective pair potential $\Gamma_S$
is the coupling strength between the dot and the superconducting electrode,
and acquires the form $\Gamma_S=2\pi|V^{S}|^2\rho_{S}$.
Here, $\rho_{S}$ denotes the density of states of the superconductor in the normal state,
whereas $V^S$ is the relevant tunnel amplitude between the dot and superconducting electrode.

The eigenstates of the effective dot's Hamiltonian (\ref{Eq:5}) can be easily found to be:
the singly occupied dot $\ket{\sigma}$, with either spin-up or spin-down,
and the two states being the superpositions of the empty $|0\rangle$
and doubly occupied $|2\rangle$ states:
\begin{equation}\label{Eq:5a}
  |\pm\rangle=\frac{1}{\sqrt{2}}\left(\sqrt{1\mp\frac{\delta}{2\varepsilon_A}}\;|0\rangle
  \mp\sqrt{1\pm\frac{\delta}{2\varepsilon_A}}\;|2\rangle\right).
\end{equation}
The corresponding eigenenergies are:
$E_\uparrow = E_\downarrow = \varepsilon$,
and $E_{\pm}=\delta/2\pm\varepsilon_A$,
where $\delta=2 \varepsilon+U$
denotes the detuning from the particle-hole symmetry point,
whereas $\varepsilon_A = \sqrt{\delta^2+\Gamma_S^2}\;/2$ measures
the energy difference between the states $|+\rangle$ and $|-\rangle$.

The Andreev bound state energies can be defined as excitation energies
of the effective dot Hamiltonian~\cite{konigPRB10}
\begin{equation}\label{Eq:6}
E_{\alpha\beta}^A=\alpha\frac{U}{2}+\frac{\beta}{2}\sqrt{\delta^2+\Gamma_S^2},
\end{equation}
where $\alpha,\beta=\pm$. These energies are defined by
differences between the eigenenergies of the dot
decoupled from the ferromagnetic leads.
$E_{++}$ ($E_{+-}$) denotes the excitation 
from the singly occupied state $\ket{\sigma}$ to state $\ket{+}$ ($\ket{-}$),
while the excitation energies $E_{--}$  and $E_{-+}$ are related with the opposite transitions.

We would like to note that by using the effective Hamiltonian (\ref{Eq:5}),
we assumed that $\Delta$ is the largest energy scale in the problem,
which implies that $\Delta > U$. Clearly, this condition may not be fulfilled in 
any hybrid device with a quantum dot coupled to superconducting lead.
However, there are superconductors,~\cite{nagamatsu01,heinrich13}
in which the energy gap can be as large as a couple of meV,
while the charging energy in quantum dots depends on their size
and can be made arbitrarily small. Here, however, one needs to balance between
the charging and thermal energies, so that the finite size and charging effects
are not smeared out by thermal fluctuations.
Consequently, the experimental implementation of hybrid quantum dots 
where the condition $\Delta > U$ is fulfilled is feasible.


\subsection{Calculation method}


To calculate the transport characteristics of the considered system
we employ the real-time diagrammatic technique,~\cite{schoeler}
adopted for the case with superconducting lead.~\cite{palaNJP07,governalePRB08}
The technique is based on the perturbation expansion of the reduced
density matrix and the relevant operators with respect
to the coupling strength to ferromagnetic leads.
After integrating out the non-interacting electronic degrees of freedom
in the ferromagnetic leads, the system is described by
reduced density matrix $\hat{\rho}$,
which in the steady-state is governed by the following equation~\cite{schoeler}
\begin{equation}\label{Eq:7}
\sum_{\chi'}\Sigma_{\chi,\chi'}P_{\chi'}=0,
\end{equation}
where $P_{\chi}$ denotes an element of the reduced density matrix,
$P_{\chi}=\langle\chi|\hat{\rho}|\chi\rangle$,
taken in the dot's state $|\chi\rangle$,
whereas $\Sigma_{\chi,\chi'}$ are the self-energies corresponding
to the evolution forward in time from state
$|\chi'\rangle$ to state $|\chi\rangle$, and then
backward in time from state $|\chi\rangle$ to state $|\chi'\rangle$.
Note that Eq.~(\ref{Eq:7}) describes only diagonal elements
of the reduced density matrix.
Generally, off-diagonal elements corresponding to coherent superpositions
of the states $|+\rangle$ and $|-\rangle$ should be also considered.
However, in the case of $\Gamma_\beta\ll\Gamma_S$ ($\beta=L,R$),
as considered in this paper,
the transition rates between states $|+\rangle$ and $|-\rangle$
become irrelevant and can be thus neglected.
\cite{konigPRB10}
The diagonal elements of the reduced density matrix,
$P_{\chi}$, simply denote the probability of finding the dot
(in the proximity with superconductor) in the state $|\chi\rangle$.
The current flowing from ferromagnetic $\beta$-lead
can be calculated from the following formula~\cite{schoeler}
\begin{equation}\label{Eq:8}
  I_{\beta}=-\frac{ie}{\hbar}\sum_{\chi,\chi'}\Sigma_{\chi,\chi'}^{I_\beta}P_{\chi'}
\end{equation}
for $\beta=L,R$, which can be also expanded order by order in tunneling processes.
Here, $\Sigma_{\chi,\chi'}^{I_\beta}$ denotes the generalized self-energy,
which takes into account the number of electrons transferred through given junction $\beta$.

In the weak coupling regime, $\Gamma_\beta\ll k_BT$,
Eq.~(\ref{Eq:7}) can be solved order by order
in the perturbation expansion in the coupling strength to ferromagnetic leads.
Each term of expansion can be visualized graphically
as a diagram (or sum of diagrams) defined on the
Keldysh contour, where the vertices are connected by lines corresponding
to tunneling processes. The self-energies in respective
order of expansion can be calculated using the respective diagrammatic rules.~\cite{konigPRB10,weymannPRB08}
Having determined the respective self-energies and occupation probabilities,
the current can be then calculated in given order of expansion.
Here, we have calculated the relevant contributions up
to the second order of expansion.~\cite{weymannPRB08}
The first order of expansion corresponds to
sequential tunneling processes, which dominate the current for
voltages larger than a threshold voltage.
Below the threshold, however, the sequential tunneling is exponentially suppressed
due to the charging energy and the system is in the Coulomb blockade regime.
In the Coulomb blockade regime transport is dominated
by cotunneling processes, which involve two correlated-in-time single
tunneling events occurring through virtual states of the
system.~\cite{averin}
These processes are captured by the second order of expansion.
Note that in the hybrid device considered here, these
processes are also related with transferring Cooper pairs between
the ferromagnets and superconductor.

Due to the proximity of superconducting lead and
Andreev processes between the dot and superconductor,
the currents flowing through the left and right junctions are generally not equal, $I_L \neq I_R$.
The total current flowing into the superconductor can be thus simply obtained from the Kirchhoff's law
\begin{equation}\label{Eq:9}
  I_{S} = I_L + I_R \,.
\end{equation}


\section{Results and discussion}


In this section we present the numerical results for Andreev transport
through quantum dot in the limit of large superconducting gap.
Particularly, we show the charge current injected (extracted) into
(from) superconducting lead and the corresponding differential conductance.
We also calculate the TMR effect associated with the change of magnetic configuration
of the ferromagnetic leads from the parallel into antiparallel alignment.
The latter quantity is defined as the ratio
\be\label{Eq:10}
   {\rm TMR} = \frac{I_S^{\rm AP}-I_S^{\rm P}}{I_S^{\rm P}},
\ee
where $I_S^{\rm P}$ and $I_S^{\rm AP}$ denote the Andreev
current flowing into superconductor in the parallel
and antiparallel magnetic configurations, respectively.

For the three-terminal setup considered in this paper,
the TMR can be used to quantify the role of crossed
Andreev reflection, as compared to direct Andreev reflection (DAR).~\cite{konig09}
The rate of a direct Andreev process is proportional to a product
of the coupling constants for spin-up and spin-down of the same junction.
This is contrary to crossed Andreev reflection,
the rate of which is proportional to the product of respective couplings, but for different leads.
Thus, any change in magnetic configuration leads to a change in CAR,
while direct Andreev processes are not affected.
Consequently, tunnel magnetoresistance provides the relevant,
though indirect, information about crossed Andreev reflection in the system,
since any change in TMR is related with a change of CAR.
In an extreme situation when the leads become half-metallic,
there are only crossed-Andreev processes possible. The current
is then maximized in the antiparallel configuration and completely suppressed
in the parallel one.

In the numerical analysis we assume that the system is symmetric,
i.e. $\Gamma_L=\Gamma_R=\Gamma/2$ and $p_L=p_R=p$.
We also assume that the voltage drop is applied symmetrically between the 
magnetic leads and superconductor, $\mu_L=\mu_R=eV$ and $\mu_S=0$.
For this configuration of applied voltages, the effect of the left-right contact asymmetry
is rather intuitive, as it mainly leads to quantitative changes.
For example, increasing the spin polarization of one lead,
generally boosts the TMR by a certain factor, while its bias and gate voltage dependence
qualitatively almost do not change.
In our analysis we also assume that the external magnetic field $B_z$ required to switch the
magnetic configuration of the system from parallel to antiparallel one and vice versa
is so small that it does not lead to the splitting of the dot level, i.e., $B_z\ll\Gamma$.


\subsection{Andreev current and differential conductance}


When applying negative bias voltage, $eV > 0$ (note that $e<0$),
one injects pairs of electrons into the superconductor,
whereas for positive bias, $eV < 0$, the Cooper pairs
are annihilated from the superconducting electrode
and entangled pairs of electrons are injected into ferromagnetic leads.
If the two electrons coming from the same lead enter
the superconductor as a Cooper pair, such process is called direct Andreev reflection.
In turn, crossed Andreev reflection occurs when the electrons
leaving from spatially separated ferromagnetic leads are injected into the superconductor.
For negative bias voltages inverse processes take place.
Particularly, a Cooper pair leaving the superconductor can tunnel
to the same ferromagnetic electrode (DAR) or split into entangled electrons
injected into separate leads (CAR).

\begin{figure}[t]
\includegraphics[width=0.8\columnwidth]{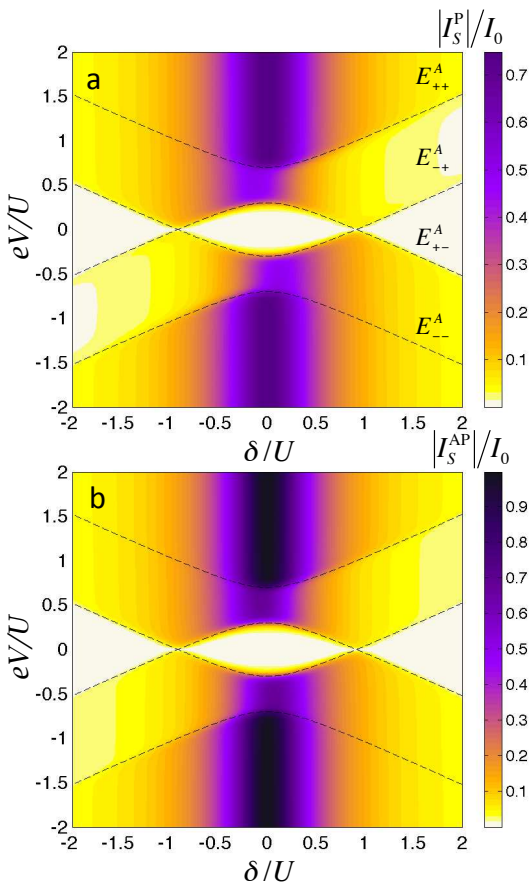}
  \caption{\label{Fig:2}
  (color online) The absolute value of the total
  Andreev current for the (a) parallel ($I^{\rm P}_S$) and (b) antiparallel ($I^{\rm AP}_S$)
  magnetic configuration as a function of detuning $\delta = 2\e+U$
  and applied bias voltage $V$.
  The dashed lines indicate the position
  of the respective Andreev bound states, see \eq{Eq:6}, as denoted in (a).
  The parameters are:
  $\G_S=0.4$, $\G=0.01$, $T=0.02$, with $U\equiv 1$ the energy unit, and $p=0.5$.
  The current is plotted in units of $I_0 = e\Gamma/\hbar$.
  The same color scale is used in (a) and (b) to enable direct comparison.}
\end{figure}

In the considered system the current flowing for negative electrochemical potential shift, $eV<0$,
is related to the current flowing for $eV>0$ by making the transformation
$eV\rightarrow -eV$, $\delta\rightarrow -\delta$, and $I_S\to -I_S$.
Thus, it is generally sufficient to consider the case of only one bias polarization.
Before analyzing the magnetoresistive properties of our hybrid device,
let us first discuss the general behavior of the Andreev current depending on the transport region.
The bias voltage $eV$ and detuning $\delta=2\varepsilon+U$ dependence of the absolute value
of the total current flowing between the superconducting and ferromagnetic leads
is shown in \fig{Fig:2} for the parallel and antiparallel magnetic configurations.
First of all, one can see that the Andreev current becomes
significant only for small values of the detuning parameter $\delta$.
This is due to the fact that the Andreev tunneling
is optimized when the particle-hole symmetry holds, i.e. in the case of $\delta \to 0$.
Furthermore, the Andreev current vanishes for small bias voltages,
$E^A_{-+} < eV < E^A_{+-}$, and for $|\delta|<\sqrt{U^2-\Gamma_S^2}$,
when the dot is occupied by a single electron.
As there are two electrons required to form a Cooper pair,
no current is flowing into the superconductor.
For $|\delta|=\sqrt{U^2-\Gamma_S^2}$, i.e. the point in which the two levels
$E^A_{+-}$ and $E^A_{-+}$ cross each other, see Fig.~\ref{Fig:2}(a),
the Andreev current vanishes only for zero bias voltage.
In turn, for $|\delta|>\sqrt{U^2-\Gamma_S^2}$, the current is
suppressed for a finite region of low bias voltage, namely 
that, which corresponds to empty or doubly occupied dot regime.
The Andreev current starts to flow when the chemical potentials
of ferromagnetic leads cross the Andreev bound state energies $E^A_{+-}$ or $E^A_{-+}$.
The current grows again when the next Andreev levels
$E^A_{++}$ or $E^A_{--}$ enter the bias window, see \fig{Fig:2}.

\begin{figure}[t]
\includegraphics[width=0.81\columnwidth]{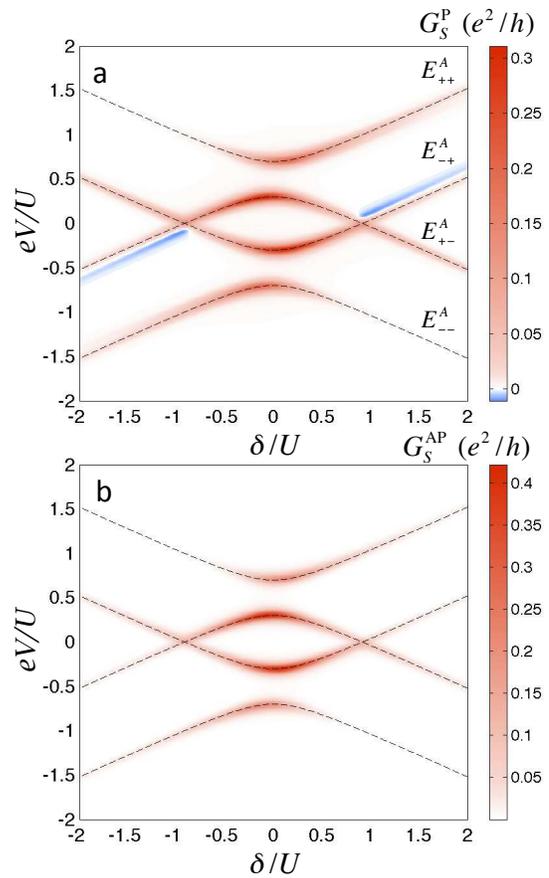}
  \caption{\label{Fig:3}
  (color online) The total differential conductance $G_S = dI_S/dV$
  due to the Andreev current for the (a) parallel ($G^{\rm P}_S$)
  and (b) antiparallel ($G^{\rm AP}_S$)
  magnetic configuration as a function of detuning $\delta = 2\e+U$
  and applied bias voltage $V$.
  The dashed lines indicate the position
  of the respective Andreev bound states.
  The parameters are as in \fig{Fig:2}.}
\end{figure}

The above described behavior is also clearly visible in the
dependence of the differential conductance on the bias voltage $eV$
and the detuning parameter $\delta$, which is shown in Fig.~\ref{Fig:3}.
Each time the electrochemical potential of ferromagnetic leads $eV$
meets the Andreev level, there appears a peak in the differential conductance.
Generally, the bias voltage dependence of the differential conductance
reveals four peaks, except for the values of detuning $\delta = \pm \sqrt{U^2-\G_S^2}$,
for which the states $E_{+-}^A$ and $E_{-+}^A$ become degenerate,
see Fig.~\ref{Fig:3} for $\delta \approx \pm U$.
Furthermore, the behavior of the transport characteristics is asymmetric
with respect to the bias reversal, which is nicely visible in the differential conductance.
More specifically, the magnitude of the peak associated with the Andreev
level $E^A_{++}$ for $eV>0$ and $\delta>0$
is larger than the amplitude of the corresponding
level $E^A_{--}$ for $eV<0$ (and $\delta>0$).
A similar asymmetry can be observed for the same Andreev states,
but for opposite detuning, $\delta<0$.
The reason for this asymmetric behavior
of the Andreev current can be understood as follows.
Let us consider the asymmetry occurring for large positive values of $\delta$.
In this case the energies $\varepsilon$ and $\varepsilon+U$
lie above the chemical potential of the superconducting lead and the dot is empty.
When shifting the electrochemical potentials of ferromagnetic leads down, $eV<0$,
there are no states in the transport window and the tunneling rate
of Cooper pairs into superconductor becomes strongly suppressed.
However, it is sufficient to change either the sign of the bias voltage $V$
or the sign of detuning $\delta$ to allow for Andreev tunneling.
This leads to highly asymmetric dependence of the differential conductance 
on the bias voltage and detuning, which is visible in both the parallel
and antiparallel magnetic configurations of the device, see Fig.~\ref{Fig:3}.

The Andreev current flows due to both direct and crossed Andreev processes.
To quantify the role of these processes, one can use ferromagnetic contacts
and, by introducing the spin-dependence of tunneling processes,
study how the current and differential conductance depend on the magnetic configuration
of the device.~\cite{konig09} The presence of CAR reveals itself just
in the dependence of the Andreev current on the magnetic configuration.
More specifically, in the antiparallel configuration
both electrons forming a Cooper pair
belong to either the majority or minority bands of the ferromagnets. 
On the other hand, in the parallel configuration one of the electrons is always a majority one, while
the other one is the minority one. The conductance is then determined by
the minority spin band of the ferromagnet.
The current flowing in the parallel magnetic configuration
is therefore suppressed as compared to the current flowing in the antiparallel configuration.
This can be clearly seen in the bias voltage and detuning dependence
of both the current, see \fig{Fig:2}, and the differential conductance, see \fig{Fig:3}.
Note that this behavior is just opposite to the typical
quantum dot spin valves when usually the conductance in the parallel
configuration is larger than that in the antiparallel configuration.
\cite{rudzinski,braun,weymann}

Moreover, the differential conductance calculated for the parallel
alignment reveals another interesting feature.
For $|\delta|>\sqrt{U^2-\Gamma_S^2}$, some negative values
of the differential conductance appear in the vicinity
of $E_{-+}^A$ ($E_{+-}^A$) for positive (negative)
electrochemical potential shift, see Fig.~\ref{Fig:3}(a).
This can be understood by realizing that in the parallel configuration
there are more spin-up electrons than spin-down ones
at the Fermi level of ferromagnetic leads.
Since injecting (or extracting) Cooper pair to the superconductor
involves two spins of opposite directions, 
the rate of electron pairs is mainly determined by the density of states of minority carriers,
which is the bottleneck for Andreev transport in the parallel configuration.
Such spin imbalance leads to the nonequilibrium spin accumulation in the dot
that develops when the dot occupancy is odd.
The occupation of the spin-up level becomes then greatly enhanced
as compared to the spin-down one (the situation can be reversed by applying opposite bias voltage),
suppressing the Andreev current and giving rise to negative differential conductance.
When further increasing the bias voltage, the occupancy of the dot changes to even
(for $|eV|\approx E_{++}^A$ or $|eV|\approx E_{--}^A$) and the dependence on spin
is weakened. Consequently, no negative differential conductance occurs then.
On the other hand, for $|\delta|<\sqrt{U^2-\Gamma_S^2}$
and low bias voltage, the dot is in the Coulomb blockade regime
and the Andreev current is generally suppressed. When reaching the threshold voltage,
the singlet states of the dot start participating in transport and the current changes monotonically.
We also note that for the antiparallel magnetic configuration
no negative differential conductance occurs, since in this case
both electrons forming a Cooper pair belong to
the same subband of the ferromagnets and there is no spin accumulation.


\subsection{Tunnel magnetoresistance}


\begin{figure}[t]
\includegraphics[width=0.8\columnwidth]{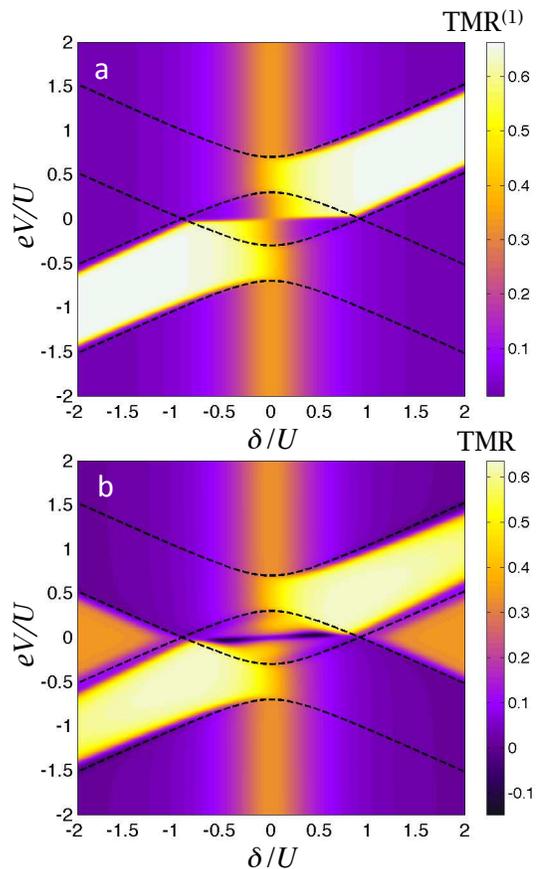}
  \caption{\label{Fig:4}
  (color online) The tunnel magnetoresistance (TMR)
  of the Andreev current as a function of detuning $\delta$ and the bias voltage $V$
  calculated by using only sequential tunneling processes [${\rm TMR}^{(1)}$] (a)
  and the first and second-order tunneling processes (b).
  The parameters are the same as in \fig{Fig:2}.
  The dashed lines show the position of Andreev bound states.
  The same color scale is used in (a) and (b) to facilitate comparison.}
\end{figure}

\begin{figure}[t]
\includegraphics[width=0.82\columnwidth]{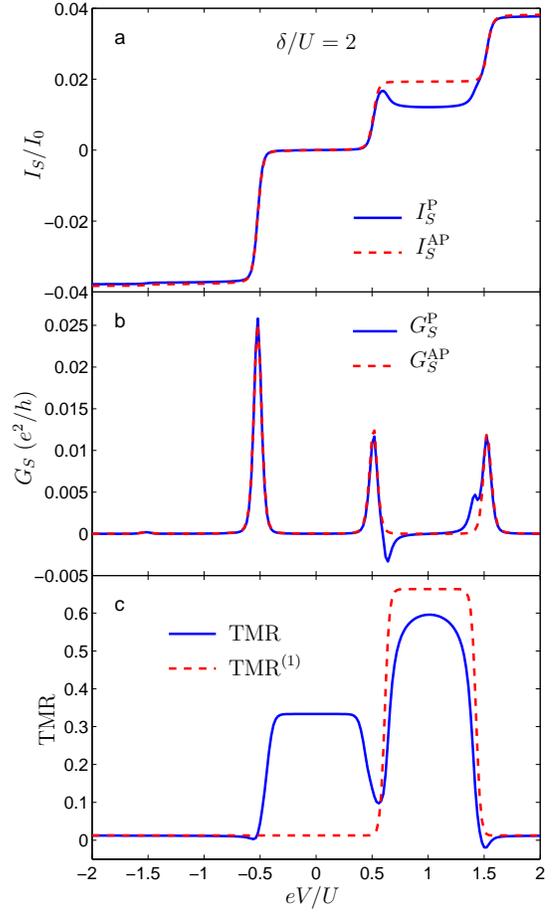}
  \caption{\label{Fig:5}
  (color online) The Andreev current (a)
  in the parallel (solid line) and antiparallel (dashed line)
  magnetic configurations, the corresponding differential
  conductance (b) and the tunnel magnetoresistance (c).
  For comparison in panel (c) we also show the TMR
  calculated using only sequential tunneling processes (dashed line).
  The parameters are the same as in \fig{Fig:2}.
  }
\end{figure}

\begin{figure}[t]
\includegraphics[width=0.8\columnwidth]{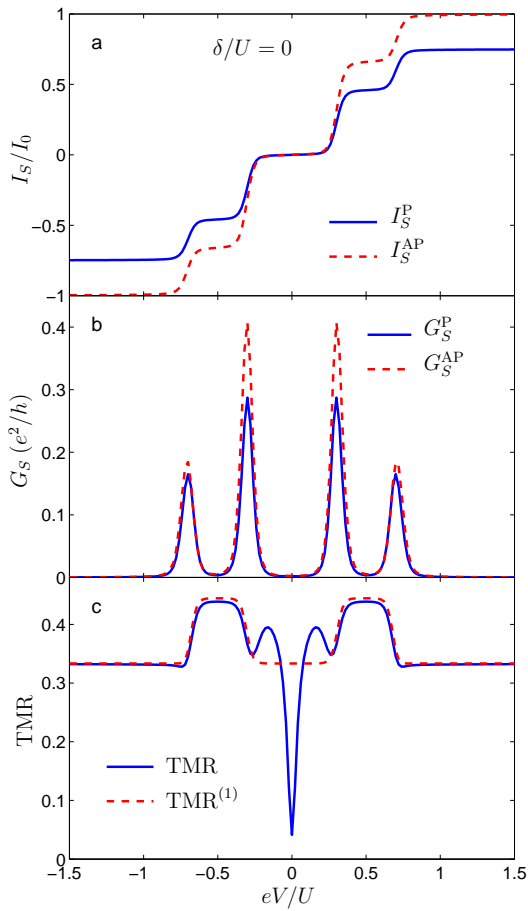}
  \caption{\label{Fig:6}
  (color online)
  The same as in \fig{Fig:5} calculated for detuning $\delta=0$.
  }
\end{figure}

As mentioned above, the presence of crossed Andreev processes
can be revealed by studying the change in transport properties
when the magnetic configuration of the device is varied.
In this section we therefore analyze the behavior of the tunnel magnetoresistance
on bias voltage and the detuning parameter $\delta$.
Figure \ref{Fig:4} presents both the total TMR as well as the TMR 
calculated using only the first-order tunneling processes [${\rm TMR}^{(1)}$].
In addition, in Figs.~\ref{Fig:5} and \ref{Fig:6} we show the
bias voltage dependence of the current, differential conductance and the TMR
for two different values of detuning $\delta$, for which the
effect of second-order processes is most visible.
It is clearly visible that cotunneling processes introduce
qualitative differences in the behavior of TMR, as compared to 
TMR obtained considering only first-order processes.
Firstly, it can be seen that the sequential TMR is positive in the whole
transport regime, see \fig{Fig:4}(a),
while the total TMR exhibits negative values for small bias voltages
and in the range of detuning parameter $|\delta|<\sqrt{U^2-\Gamma_S^2}$, see \fig{Fig:4}(b).
Secondly, for $|\delta| > \sqrt{U^2-\Gamma_S^2}$
and for bias voltages $E_{+-}^A<eV<E_{-+}^A$,
the total TMR becomes finite, while the sequential TMR is much suppressed.
To explain the behavior of the TMR on the transport regime,
let us now discuss the relevant cross-sections of the density plots shown in \fig{Fig:4}.

In Fig.~\ref{Fig:5} we plotted the bias dependence of the current (a), differential conductance (b)
for parallel and antiparallel configurations, and the TMR effect (c) for detuning $\delta/U=2$.
As described above, the voltage dependence reveals asymmetry
with respect to the bias reversal. There are three steps in the Andreev current
as a function of the bias voltage, which correspond to the respective Andreev peaks in the differential conductance,
see Figs.~\ref{Fig:5}(a) and (b). One can also notice the regime
of suppressed current in the parallel configuration visible for $U/2 \lesssim eV \lesssim 3U/2$
and the associated negative differential conductance.
The mechanism leading to this behavior was explained in previous section.
Let us now therefore focus on the behavior of the TMR effect,
which is plotted in Fig.~\ref{Fig:5}(c).
For large detuning $\delta$ there is in principle one transport regime
where the sequential tunneling approximation clearly gives wrong result for the TMR.
This is in the low bias voltage regime where the sequential processes are suppressed.
The total TMR exhibits then a plateau for
$E^A_{+-}<eV<E^A_{-+}$ ($-U/2 \lesssim eV \lesssim U/2$),
whereas the corresponding ${\rm TMR^{(1)}}$,
calculated taking into account only the first-order processes, is strongly suppressed.
When increasing the bias voltage, $E^A_{-+}<eV<E^A_{++}$ ($U/2 \lesssim eV \lesssim 3U/2$), 
the current becomes dominated by sequential tunneling, however
second-order processes still lead to a considerable modification of ${\rm TMR^{(1)}}$,
namely to its lowering. On the other hand, for 
$eV>E^A_{++}$ ($eV \gtrsim 3U/2$) or $eV<E^A_{+-}$ ($eV \lesssim -U/2$),
the effect of second-order tunneling on TMR is rather negligible.

Since tunnel magnetoresistance exhibits well-defined
transport regions where it is constant, one can derive some
approximative analytical formulas for the TMR.
This can be done assuming very low temperatures, so that 
one can replace the Fermi functions by step functions.
Then, the formula for ${\rm TMR^{(1)}}$ calculated for $\delta/U=2$
describing the value in the plateau ($U/2 \lesssim eV \lesssim 3U/2$) is given by
\be \label{eq:TMR1}
{\rm TMR^{(1)}} = \frac{2(1+\varepsilon_A)}{(1+3\varepsilon_A)}\frac{2p^2}{1-p^2},
\ee
while the sequential TMR for bias voltages $eV \gtrsim 3U/2 $ is given by
\be \label{eq:TMR2}
{\rm TMR^{(1)}} = \frac{\varepsilon_A^2-1}{2\varepsilon_A^2}\frac{2p^2}{1-p^2}.
\ee
This analytical expression also describes the TMR for $eV \lesssim -3U/2$.
On the other hand, the TMR for $-3U/2 \lesssim eV \lesssim -U/2$ reads
\be \label{eq:TMR3}
{\rm TMR^{(1)}} = \frac{2(\varepsilon_A-1)}{(3\varepsilon_A-1)}\frac{2p^2}{1-p^2}.
\ee
Note that the $\varepsilon_A$-dependent ratio in \eq{eq:TMR1} tends to unity
for $\delta = 2U$, whereas the other two factors from Eqs.~(\ref{eq:TMR2}) and (\ref{eq:TMR3})
approach zero, so that TMR then vanishes.
In the Coulomb blockade regime, $-U/2 \lesssim eV \lesssim U/2$,
the sequential TMR is clearly wrong because the current is driven by cotunneling processes.
To find the formula for TMR in the cotunneling regime, we
assume zero temperature and low bias voltage. Moreover, 
since the dot is in the singlet state, the only relevant
cotunneling processes are the spin-conserving ones (elastic processes).
Because we are interested in TMR, which is given by appropriate ratio of the currents,
it is sufficient to consider only the dependence on the couplings
to ferromagnetic leads (the denominators of cotunneling rates will cancel).
The rate of a cotunneling process is proportional to,
$\gamma_{rr'}^{\sigma\bar{\sigma}}\sim \Gamma_r^\sigma \Gamma_{r'}^{\bar{\sigma}}$,
and corresponds to transferring spin-$\s$ electron between the dot and lead $r$
and spin-$\bar{\s}$ electron between the dot and lead $r'$.
If $r=r'$, the process is related with direct Andreev tunneling, while
for $r\neq r'$ we have crossed Andreev process.
Thus, the current is generally proportional to,
$I_S \sim \sum_{r,r' = L,R} \Gamma_r^\sigma \Gamma_{r'}^{\bar{\sigma}}$,
which, by inserting proper coupling constants for given magnetic configuration, yields
\be
  I_S^{\rm P} \sim \Gamma^2 (1-p^2) {\;\;\;\;\; \rm and \;\;\;\;\;} I_S^{\rm AP} \sim \Gamma^2,
\ee
for the parallel and antiparallel configurations, respectively. The TMR is then given by
\be
 {\rm TMR}  = \frac{p^2}{1-p^2}.
\ee
This formula approximates the TMR in the Coulomb blockade regime when
the ground state is a singlet state, which for assumed spin polarization gives ${\rm TMR}=1/3$,
see \fig{Fig:5}(c).

When the detuning is absent, the Andreev current becomes maximized.
This situation is shown in \fig{Fig:6}, where the bias dependence of the current,
differential conductance and TMR is presented. First of all, one can see
that transport characteristics are now symmetric with respect to the bias reversal.
The current as a function of the bias voltage changes monotonically
giving rise to four peaks in differential conductance, see Figs.~\ref{Fig:6}(a) and (b).
On the other hand, the TMR out of the Coulomb blockade regime
takes well defined values that can be approximated by considering
sequential processes in the following way
\be
{\rm TMR}^{(1)} = \frac{2}{3}\frac{2p^2}{1-p^2},
\ee
at the plateau for voltages where the first step in the current occurs
as the voltage is increased from $V=0$, and 
\be
{\rm TMR}^{(1)} = \frac{1}{2}\frac{2p^2}{1-p^2}
\ee
in the high voltage regime, $|eV| > E_{\pm\pm}^A$.
In the Coulomb blockade regime, $E_{-+}^A<eV<E_{+-}^A$,
the dot is singly occupied and the ground state is a doublet.
Then, there are both spin-flip and non spin-flip cotunneling processes
allowed and providing a simple analytical formula for the TMR
in this transport regime is rather not possible. Moreover,
nonequilibrium spin accumulation also builds up in the dot with increasing the bias voltage.
This altogether leads to a nontrivial dependence of the TMR
on the applied bias, which is clearly different than that predicted within
the sequential tunneling approximation, see \fig{Fig:6}(c).
The total TMR has a minimum at the zero bias, and then
starts increasing with the bias voltage to drop
again just before the threshold for sequential tunneling,
giving rise to a local maximum.

Finally, we would like to note that although the analogy to the Julliere model is here
rather unjustified, since the considered system is clearly different
from a single ferromagnetic tunnel junction,
the derived analytical formulas for the TMR can be,
funnily enough, expressed in terms of ${\rm TMR^{Jull}} = 2p^2/(1-p^2)$,
which is the value characteristic of a ferromagnetic tunnel junction.~\cite{julliere}

\begin{figure}[t]
\includegraphics[width=0.9\columnwidth]{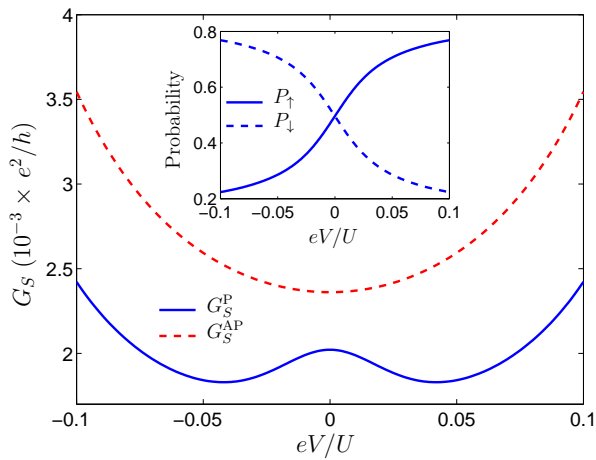}
  \caption{\label{Fig:7}
  (color online) The bias dependence of the Andreev differential conductance
  for the parallel and antiparallel magnetic configurations calculated for 
  the Coulomb blockade regime with $\delta=0$. 
  The inset shows the relevant occupation probabilities
  for spin-up and spin-down levels.
  The parameters are the same as in \fig{Fig:2}.}
\end{figure}


\subsection{Zero-bias anomaly}


\begin{figure}[t]
\includegraphics[width=0.8\columnwidth]{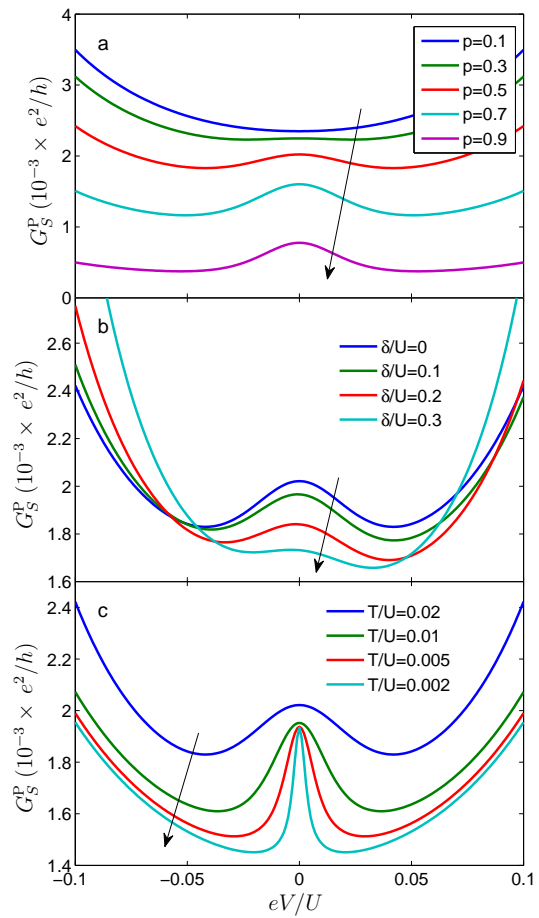}
  \caption{\label{Fig:8}
  (color online) The bias dependence of the Andreev differential conductance
  in the parallel configuration for different
  (a) spin polarization $p$ of the ferromagnets,
  (b) detuning $\delta$ and (c) temperature $T$.
  The parameters being changed are specified in each
  panel, while the other parameters are the same as in \fig{Fig:2}
  with $\delta = 0$. The arrows indicate the direction of change
  of the parameters that are tuned in each panel.}
\end{figure}

The highly nontrivial dependence of the TMR on the bias voltage in the case of 
$\delta = 0$, see \fig{Fig:6}(c), suggests that the Andreev conductance
must also reveal a nontrivial behavior. This is indeed the case.
While in the antiparallel configuration the differential conductance shows
a typical parabolic dependence on the applied bias voltage,
in the parallel configuration, on the contrary, the differential conductance
first drops when the voltage is increased, see \fig{Fig:7}.
This gives rise to a maximum in the differential
conductance in the parallel configuration at the zero bias.
The maximum bears a strong resemblance to the zero-bias anomaly
predicted for quantum dots coupled to ferromagnetic leads,
which occurs when the magnetic configuration of the device is antiparallel.~\cite{weymannZBA}
Here, in the considered hybrid device, the zero-bias anomaly develops in the
parallel configuration and is a direct consequence of the dependence
of Andreev processes on the spin polarization of ferromagnets.
As can be seen in the inset of \fig{Fig:7}, with increasing the bias voltage,
nonequilibrium spin accumulation builds up in the dot:
the occupation probability of the spin-up level is different from that of the spin-down level.
As a consequence, tunneling of Cooper pairs becomes suppressed
and the differential conductance drops. 
With further increase of the bias voltage, the rate of Andreev processes
is enhanced and the conductance starts increasing again,
despite the presence of strong nonequilibrium spin accumulation.

To discuss more specific features of the zero-bias anomaly
of the Andreev conductance in the parallel configuration,
in \fig{Fig:8} we plot $G^{\rm P}_S$ for different spin polarization
of the ferromagnets, different detuning and different temperatures.
When the spin polarization $p$ is low, see the curve for $p=0.1$
in \fig{Fig:8}(a), the maximum in $G^{\rm P}_S$ is hardly visible
and starts developing only when the spin polarization is increased, 
see the curves for $p\gtrsim 0.3$. Then, increasing $p$ leads generally to
an enhancement of the relative height of the zero-bias anomaly,
although the magnitude of the conductance is gradually decreased and
should be fully suppressed for $p\to1$, when it is not possible to
form a Cooper pair any more since the ferromagnets support only one spin species.
The enhancement of the relative height of the peak in $G^{\rm P}_S$ at zero bias
is related with the fact that by increasing $p$, the nonequilibrium spin accumulation
is also increased, which leads to a stronger suppression of the conductance
with increasing the bias voltage.

The zero-bias anomaly occurs in the Coulomb blockade regime
when the dot occupation is odd and it is most pronounced
in the case of zero detuning $\delta = 0$. When moving away
from the symmetry point, $\delta \neq 0$, the relative height of the maximum 
in $G^{\rm P}_S$ is lowered and for large enough detuning
it completely vanishes, see \fig{Fig:8}(b). This is simply related with the fact
that by increasing $|\delta|$, the Coulomb blockade is weakened 
and the role of the second-order processes is diminished as compared
to the first-order processes. In fact, the role of various second-order processes
in the formation of the zero-bias peak can be understood from the bias dependence
of $G^{\rm P}_S$ calculated for different temperatures, see \fig{Fig:8}(c).
It is clearly visible that with lowering the temperature, the width
of the zero-bias anomaly is decreased. This dependence is similar
to that observed for the zero-bias anomaly, which occurs in the antiparallel configuration
for typical quantum dot spin valves.~\cite{weymannZBA}
It indicates the role of the single-junction spin-flip cotunneling processes.
In the considered three-terminal setup, where the
electrochemical potentials of both ferromagnets are kept the same,
such spin-flip processes can occur by involving either the two ferromagnetic leads
or just a single lead. These processes do not contribute to the Andreev transport,
although the Cooper pairs can be created or annihilated in virtual states,
however they can change the dot occupations and thus indirectly affect the current.
The rate of such spin-flip processes is proportional to temperature,
$\gamma_{rr'}^{\s\bar{\s}} \sim \G_r^\s \G_{r'}^{\bar{\s}}  T \e^2 / (\e^2 - \G_S^2/4)^2$.
Consequently, with lowering $T$ the amount of spin flips
is reduced, so that the nonequilibrium spin accumulation sets in
for lower voltages, and the conductance drop is also shifted towards lower biases.
As a result, the width of the zero-bias anomaly becomes linearly reduced
with lowering the temperature.
In fact, the width of the maximum at zero bias can also provide
an information about the energy scale, at which the 
transport processes start dominating over the spin-flip processes.
Once this happens, nonequilibrium spin accumulation develops in the dot
and $G_S^{\rm P}$ first drops to further increase with raising the bias voltage.
We also note that at very low temperatures, $T\to 0$, the slip-flip processes are not present
and the zero-bias anomaly does not develop. The conductance is then due to the processes
involving the states $\ket{+}$ and $\ket{-}$, whose occupation is low but still finite,
giving rise to a finite Andreev current. Nevertheless, in the limit of a very deep Coulomb blockade,
$U\gg\G$, the current between the superconductor and ferromagnetic leads becomes fully suppressed.


\section{Conclusions}


In this paper we have analyzed the Andreev current flowing through 
a quantum dot connected to one superconducting and two ferromagnetic leads
in the sequential and cotunneling regimes. The considerations were based on the
real-time diagrammatic technique, which allowed us to systematically
study the effect of cotunneling processes on the Andreev current,
differential conductance and the resulting TMR.
The Andreev current occurs in such three-terminal setup
due to both direct and crossed Andreev reflections.
Since by changing the magnetic configuration of the device,
it is CAR which is affected, the role of crossed Andreev reflection
in transport can be quantified by studying the behavior of the TMR.

We showed that depending on the transport regime, the TMR
can take well-defined values and can be described by simple analytical formulas.
Moreover, we also showed that the cotunneling processes considerably modify
the behavior of TMR in the blockade regions where sequential processes are suppressed.
In the blockade regime where the dot ground state is a singlet,
the TMR exhibits a plateau with the value given by $p^2/(1-p^2)$,
which is in stark contrast to the highly suppressed TMR obtained within
the sequential tunneling approximation. On the other hand,
in the Coulomb blockade with single electron,
the TMR exhibits a highly nontrivial dependence
on the bias voltage with a minimum at the zero bias,
opposite to the sequential TMR, which is constant.
This nontrivial dependence results from a particular dependence
of the Andreev current and differential conductance on the applied bias voltage.
While in the antiparallel configuration the differential conductance
displays a parabolic dependence on the applied bias, in the parallel configuration
the conductance exhibits a peak at the zero bias.
The zero-bias anomaly of the Andreev conductance is related
with a nonequilibrium spin accumulation on the dot
and a subtle interplay between the spin-flip cotunneling processes
that do not contribute to the Andreev current and the processes
that drive the current. Since the rate of the former processes depends
on temperature, while that of the latter processes on the applied bias voltage,
the width of the zero-bias anomaly strongly depends on temperature
and is decreased with lowering $T$.

\section*{Acknowledgments}

We acknowledge support from the `Iuventus Plus' project No. IP2011 059471 for years 2012-2014.
I.W. also acknowledges support from the EU grant No. CIG-303 689.
P.T. also acknowledges support by the European Union under
European Social Fund Operational Programme
`Human Capital' (POKL.04.01.01-00-133/09-00).


\end{document}